\documentclass[twocolumn]{bmcart}% uncomment this for twocolumn layout and comment line below
\usepackage{amsthm,amsmath,amssymb,bbm}
\usepackage[utf8]{inputenc} %unicode support
\usepackage{tabularx}
\usepackage{hhline}
\usepackage{tikz}
\usepackage{mathtools}
\usepackage{cuted}
\usepackage{graphicx}
\usepackage{listings,lstautogobble}
\usepackage{xcolor}

\usepackage{subfigure}

\allowdisplaybreaks

\definecolor{RoyalBlue}{cmyk}{1, 0.50, 0, 0}

\startlocaldefs

\endlocaldefs

\begin{document}
	
	%%% Start of article front matter
	\begin{frontmatter}
		
		\begin{fmbox}
			\dochead{Review}
			
			
			\title{Next Generation Backscatter Communication: {Systems, Techniques and Applications}}
			
			%%%%%%%%%%%%%%%%%%%%%%%%%%%%%%%%%%%%%%%%%%%%%%
			%%                                          %%
			%% Enter the authors here                   %%
			%%                                          %%
			%% Specify information, if available,       %%
			%% in the form:                             %%
			%%   <key>={<id1>,<id2>}                    %%
			%%   <key>=                                 %%
			%% Comment or delete the keys which are     %%
			%% not used. Repeat \author command as much %%
			%% as required.                             %%
			%%                                          %%
			%%%%%%%%%%%%%%%%%%%%%%%%%%%%%%%%%%%%%%%%%%%%%%
			
			\author[
			addressref={aff1},                   % id's of addresses, e.g. {aff1,aff2}
			%corref={aff1},                       % id of corresponding address, if any
			%noteref={n1},                        % id's of article notes, if any
			email={wanchun.liu@sydney.edu.au}   % email address
			]{\inits{WL}\fnm{Wanchun} \snm{Liu}}
			\author[
			addressref={aff2},
			corref={aff2},
			email={huangkb@eee.hku.hk}
			]{\inits{KH}\fnm{Kaibin} \snm{Huang}}
			\author[
			addressref={aff3},
			email={}
			]{\inits{XZ}\fnm{Xiangyun} \snm{Zhou}}
			\author[
			addressref={aff3},
			email={}
			]{\inits{SD}\fnm{Salman} \snm{Durrani}}
			
			%%%%%%%%%%%%%%%%%%%%%%%%%%%%%%%%%%%%%%%%%%%%%%
			%%                                          %%
			%% Enter the authors' addresses here        %%
			%%                                          %%
			%% Repeat \address commands as much as      %%
			%% required.                                %%
			%%                                          %%
			%%%%%%%%%%%%%%%%%%%%%%%%%%%%%%%%%%%%%%%%%%%%%%
			
			\address[id=aff1]{%                           % unique id
				\orgname{School of Electrical and Information Engineering, The University of Sydney}, % university, etc
				%\street{},                     %
				\postcode{2006}                                % post or zip code
				\city{Sydney, NSW},                              % city
				\cny{Australia}                                    % country
			}
			\address[id=aff2]{%
				\orgname{Department of Electrical and Electronic Engineering, The University of Hong Kong},
				%\street{},
				%\postcode{310027}
				\city{Hong Kong},
				\cny{China}
			}
			\address[id=aff3]{%
				\orgname{Research School of Engineering, The Australian National University}, % university, etc
				%\street{},                     %
				\postcode{2601}                                % post or zip code
				\city{Canberra, ACT},                              % city
				\cny{Australia}
			}
			
			%%%%%%%%%%%%%%%%%%%%%%%%%%%%%%%%%%%%%%%%%%%%%%
			%%                                          %%
			%% Enter short notes here                   %%
			%%                                          %%
			%% Short notes will be after addresses      %%
			%% on first page.                           %%
			%%                                          %%
			%%%%%%%%%%%%%%%%%%%%%%%%%%%%%%%%%%%%%%%%%%%%%%
			
			\begin{artnotes}
				%\note{Sample of title note}     % note to the article
				%\note[id=n1]{Equal contributor} % note, connected to author
			\end{artnotes}
			
			%\end{fmbox}% comment this for two column layout
			
			%%%%%%%%%%%%%%%%%%%%%%%%%%%%%%%%%%%%%%%%%%%%%%
			%%                                          %%
			%% The Abstract begins here                 %%
			%%                                          %%
			%% Please refer to the Instructions for     %%
			%% authors on http://www.biomedcentral.com  %%
			%% and include the section headings         %%
			%% accordingly for your article type.       %%
			%%                                          %%
			%%%%%%%%%%%%%%%%%%%%%%%%%%%%%%%%%%%%%%%%%%%%%%
			
			\begin{abstractbox}
				
				\begin{abstract} % abstract
					The rapid growth of IoT driven by recent advancements in consumer electronics, 5G communication technologies, and cloud-computing enabled big-data analytics, has recently attracted tremendous attention from both the industry and academia. One of the major open challenges for IoT is the limited network lifetime due to  massive IoT devices being powered by batteries with finite capacities. The low-power and low-complexity backscatter communications (BackCom), which simply relies on passive reflection and modulation of an incident radio-frequency (RF) wave, has emerged to be a promising technology for tackling this challenge. However, the contemporary BackCom has several major limitations, such as short transmission range, low data rate and uni-directional information transmission. The article aims at introducing the recent advances in the active area of BackCom. Specifically, we provide a systematic introduction of the next generation BackCom covering basic principles, systems, techniques besides IoT applications. Lastly, we describe the IoT application scenarios with the next generation BackCom.
				\end{abstract}
				
				%%%%%%%%%%%%%%%%%%%%%%%%%%%%%%%%%%%%%%%%%%%%%%
				%%                                          %%
				%% The keywords begin here                  %%
				%%                                          %%
				%% Put each keyword in separate \kwd{}.     %%
				%%                                          %%
				%%%%%%%%%%%%%%%%%%%%%%%%%%%%%%%%%%%%%%%%%%%%%%
				
				\begin{keyword}
					\kwd{Backscatter communication}
					\kwd{IoT}
					\kwd{wireless power transfer}
					\kwd{wirelessly powered network}
					\kwd{wireless sensor network}
				\end{keyword}
				
				% MSC classifications codes, if any
				%\begin{keyword}[class=AMS]
				%\kwd[Primary ]{}
				%\kwd{}
				%\kwd[; secondary ]{}
				%\end{keyword}
				
			\end{abstractbox}
		\end{fmbox}% uncomment this for twcolumn layout
		
	\end{frontmatter}

\section{Introduction}
{In the past decades, the IoT has  seen technological innovations in a wide range of applications such as smart city, smart home, autonomous robots, vehicles and {unmanned aerial vehicles} (UAVs). {\emph{The IoT is expected to comprise tens of billions of sensors in the near future.
Keeping the massive number of  energy-constrained IoT sensors alive poses a key design challenge for IoT. This is especially challenging given a large number of the sensors may be hidden (e.g., in the walls or appliances) or deployed in remote  or hazardous environments (e.g., in radioactive areas or pressurized pipes), making battery recharging or replacement difficult if not impossible. Thus, it is highly desirable to power IoT nodes by  ambient energy harvesting~\cite{ambient_EH,Seattle,Tokyo} or wireless {power transfer} (PT)~\cite{PowerCast,Huang15}.}} One particular promising solution in this regard is {backscatter communications} (BackCom), which allows an IoT node to transmit data by reflecting and modulating an  incident RF wave~\cite{Boyer14}. The conventional  radio architecture comprises power-hungry RF chains  having  oscillators, mixers and digital-to-analog converters, which results in non-compact form factors and limits the battery life of IoT devices.
{For example, the well-adopted ZigBee CC2520 chip from Texas Instruments consumes about $100$~mW for transmission~{\cite{CC2520}}, which is quite a large power consumption.}
In contrast, a backscatter node has no active RF components and as a result can be made to have miniature hardware with extremely low power consumption, e.g., in the order of $10\ \mu$W~\cite{Tagconsumption}, facilitating large-scale deployment at a flexible location or even in-body implantation.

In the past two decades, point-to-point BackCom has been widely deployed  in the application of {radio-frequency identification} (RFID) for a passive RFID Tag to report an ID to an enquiring Reader over the near field (typically several centimeters). In its early stage, IoT comprised of primarily RFID devices for logistics and inventory management. However,  IoT is expected to connect tens of billions of devices and accomplish much more sophisticated and versatile tasks with city-wide  or even global-scale influences.
This demands the communication capabilities and ranges (tens of meters) between IoT nodes to be way beyond the primitive RFID operations supporting bursty and low-rate (several-bytes pre-written ID sequence)  uni-directional transmission over several meters. This can be achieved via a full-fledged BackCom theory leveraging the advanced communication technologies such as small-cell networks, full-duplexing\footnote{{Note that the BackCom full-duplexing is different from that of the conventional communication systems, and there is a performance tradeoff between the transmission and the reception of a full-duplex BackCom node, which will be discussed in Sec. 4.2.}}, multi-antenna communications, massive access and wireless PT, as well as advancements in electronics such as miniature radios  (e.g., button-size radios)  and low-power electronics.
Therefore, the developing IoT applications present many promising research opportunities, resulting  in a recent surge in research interests in BackCom.
Table~\ref{tab:table} summarizes and compares the important properties of the conventional RFID and the next generation BackCom systems.
\begin{table*}[t]
	\small
	\caption{{Comparison of Conventional and Next Generation BackCom Systems.}}
	\label{tab:table}

	\hspace{-1.5cm}
	\begin{tabular}{ l | c  | c| c| c| c }
		\hline			
		& Distance & Data Rate  & Modulation & Networking & Energy Source  \\
		\hline
		\hline
		Conventional RFID & $< 1$ m & $< 640$ Kbps  & Binary & Point-to-point & Dedicated  \\
		\hline
		\hspace{-0.22cm}\begin{tabular}{l}
			Next Generation \\BackCom
		\end{tabular} & $< 3$ km & $< 10$ Mbps  & \begin{tabular}{l}
		Higher-order\\ modulation
	\end{tabular} & \begin{tabular}{l}
	Multiple access \\and ad hoc
\end{tabular} & Dedicated and ambient  \\
		\hline
	\end{tabular}
\end{table*}
}

This article aims at introducing the recent advances in the active area of next generation BackCom.  First, we summarize the basic principles, system and network architectures for BackCom.  Second, we discuss how specific communication techniques have been redesigned for BackCom.
Last, we focus on the applications of BackCom for IoT.

\section{BackCom Basic Principles and Design Tradeoffs} \label{sec:basic}

\subsection{Architecture for BackCom}
\begin{figure}
	\centering
	\resizebox{0.4\textwidth}{!}{%
		\includegraphics[scale=0.52]{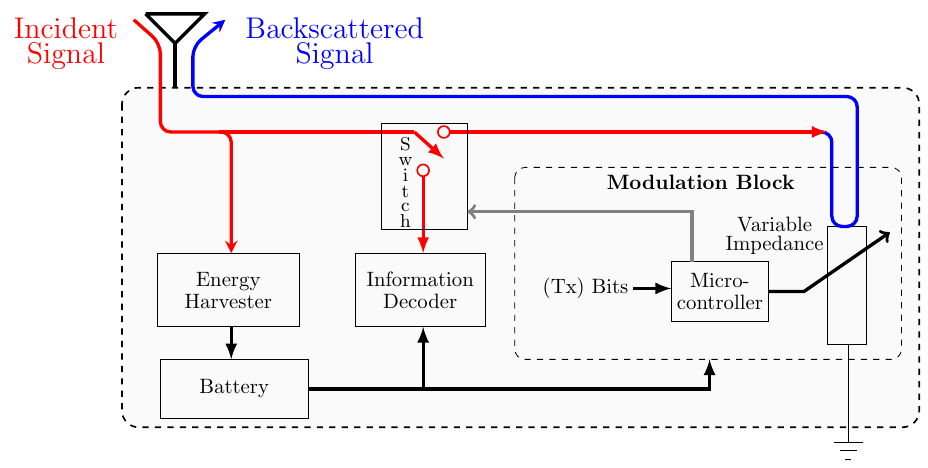}	
	}%
	\caption{{The architecture of a backscatter Tag.} }
	\label{fig:architecture}
\end{figure}

{A basic BackCom system consists of two devices: a mobile backscatter node, i.e., a Tag, and a Reader\footnote{{BackCom Readers are typically connected to the power grid or equipped with large capacity batteries.}}~\cite{Boyer14}.
The architecture of the Tag consists of an RF energy harvesting block, a battery, a modulation block and an information decoder, as illustrated in Fig.~\ref{fig:architecture}~\cite{Boyer14}\cite{WanchunBackCom}.}	
The Tag is a passive node that harvests energy from an incident single-tone sinusoidal {continuous wave} (CW) radiated by the Reader, and also modulates and reflects a fraction of the wave back to the Reader.
Specifically, the wave reflection is due to an intentional mismatch between the antenna and load impedance.
Varying the load impedance makes the reflection coefficient to vary following a random sequence that modulates the reflected wave with Tag's information bits.
Such a modulation scheme is named as the \emph{backscatter modulation}.
The passive Tag is powered by RF energy harvesting and does not require any active RF component.
On the contrary, the Reader has its own power supply and a full set of conventional RF components for emitting CW and information transmission/reception.

\begin{figure}
	\centering
	\resizebox{0.4\textwidth}{!}{%
		\includegraphics[scale=0.82]{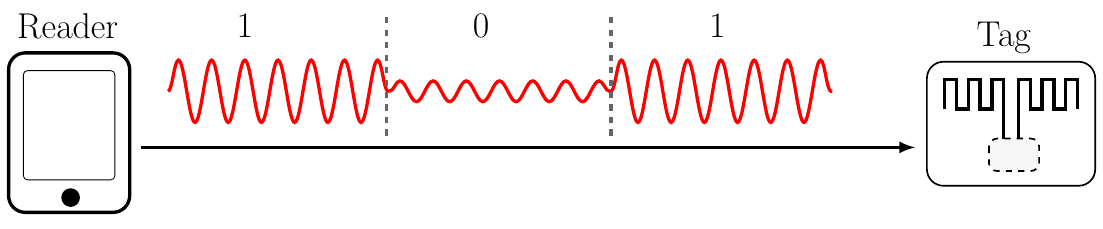}	
	}%
	\caption{{The forward information-transmission mode of a BackCom link.}}
	\label{fig:forwardIT_1}
\end{figure}

\begin{figure}
	\centering
	\resizebox{0.4\textwidth}{!}{%
		\includegraphics[scale=0.82]{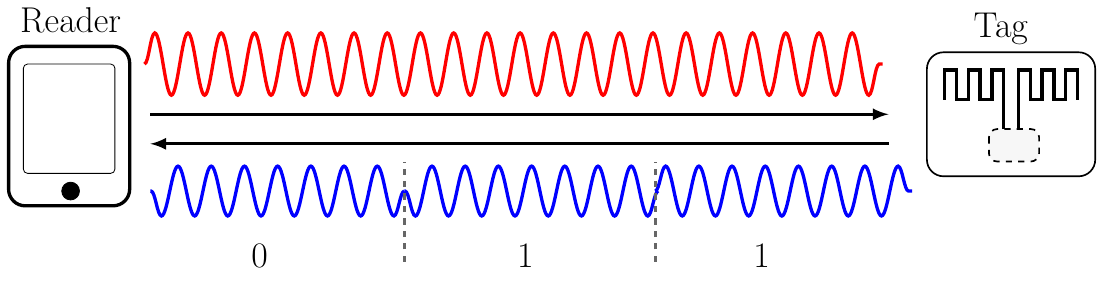}	
	}%
	\caption{{The backward information-transmission mode of a BackCom link.}}
	\label{fig:forwardIT_2}
\end{figure}

%
%\begin{figure}[t]	
%	\small
%	\renewcommand{\captionlabeldelim}{ }
%	\renewcommand{\captionfont}{\small} \renewcommand{\captionlabelfont}{\small}
%	\centering
%	\includegraphics[scale=0.6]{/Architecture}	
%	%		\vspace{-5pt}		
%	\caption{The architecture of a backscatter Tag. }
%	\label{fig:architecture}	
%	\vspace{-0.5cm}
%\end{figure}

\subsection{BackCom Modes and Modulation}
In general, the communication between the Reader and the Tag has two modes: the forward information transmission, i.e., the Reader-to-Tag transmission, and the backward information transmission, i.e., the Tag-to-Reader transmission.

For the forward information transmission, as illustrated in Fig.~\ref{fig:forwardIT_1}, the Reader transmits a binary intensity modulated signal to the Tag. The Tag connects its information decoder and utilizes the received RF signal for RF energy harvesting and energy-detection based demodulation.
For example, the decoded bit is `$1$' or `$0$' when a high or a low signal energy is detected, respectively.
%Such on/off keying scheme is common used for modulating downlink signals from a Reader to a Tag (sensor).
The use of this primitive on/off modulation and energy detection is due to the constraint that a typical Tag is provisioned with an energy detector instead of a power hungry RF chain needed for coherent demodulation.

For the backward information transmission, the Reader sends a CW signal to the Tag,
and the Tag connects its modulation block and utilizes the received RF signal for RF energy harvesting and backscatter modulation.
% e.g., by switching between a pair of impedances for a Binary Phase-shift keying (BPSK) modulation.
Generally, a Tag can modulate the reflected signal by switching over a given set of impedances, generating a set of reflection coefficients forming a constellation.
For example, assuming {binary phase-shift keying} (BPSK) modulation, as illustrated in Fig.~\ref{fig:forwardIT_2},
the Tag has a pair of impedances corresponding to two symbols, and it chooses either one of them for backscattering depending on the value of the transmitted bit.
{In practice, the switch is usually a complementary metal-oxide-semiconductor (CMOS) switch~\cite{CMOS}, which is an active element. The switch and the set of impedances can be treated as a variable impedance. Since the switch is an active component and the impedances are passive components, the variable impedance can be regarded as a partially active component.}
%there are two impedances corresponding to two symbols and each of the symbols carries one bit of information.
%
%Since most of the existing BackCom applications, such as RFID and BackCom-based sensor networks, have asymmetric data traffics, i.e., a low-rate command through the forward IT and a high-rate information-bearing data through the backward IT, the backward IT is the dominant mode.
%However, for IoT applications
%

For backward information transmission modulation, one unique design issue is the consideration of energy-harvesting efficiency.
Specifically, it is desirable to design a modulation scheme for BackCom that reduces the reflected energy and thereby increases the harvested energy.
{In general, this objective can be achieved by shifting the constellation points on the complex plane towards the origin~\cite{Boyer14}.} This, however, may increase the detection error-rate, introducing an energy-rate tradeoff.

The backward information transmission is the dominant mode for most of the conventional RFID applications,
which have asymmetric data traffics, e.g., a low-rate command through the forward information transmission and a high-rate information-bearing data through the backward information transmission.
However, the forward and backward information transmission are equally important for future IoT applications, which require peer-to-peer network architecture and symmetric communication links between the massive number of devices.
% % % %
%For the typical applications of BackCom in IoT sensor networks, the downlink and uplink transmissions have asymmetric data rates where the low-rate downlink is for sending control signals and the relatively high-rate uplink for uploading sensing data.

%\begin{figure}[t]	
%	\small
%	\renewcommand{\captionlabeldelim}{ }
%	\renewcommand{\captionfont}{\small} \renewcommand{\captionlabelfont}{\small}
%	\centering
%	\includegraphics[scale=0.6]{/Operation}	
%	%		\vspace{-5pt}		
%	\caption{The operation modes of a BackCom link: (a) forward and (b) backward information transfer (IT). }
%	\label{fig:forwardIT}	
%	\vspace{-0.5cm}
%\end{figure}

%A key characteristic of BackCom is information transmission{double} path-loss due to the fact that the backscatter signal received at a Reader propagates through the close-loop channel cascading the downlink and uplink channels. The severe path loss can result from propagation distances in IoT that are much longer than those for  RFID applications. Thus, error-control coding is necessary for protecting the data signal against channel noise and fading. Nevertheless, coding can increases the energy  consumption of Tag circuit. Thus exists an information transmission{energy-reliability tradeoff} from the coding perspective.

\subsection{Energy-Rate Tradeoff}
Besides the said tradeoff arising from modulation design, there exists another one due to bursty transmission by IoT devices. For instance, a sensor for crowd-sensing reports data only upon receiving a query and spends the remaining time on other activities e.g., sensing and computing. Leveraging this characteristic, a backscatter Tag can be designed to periodically switch between the two modes, namely the \emph{silent} (or energy harvesting) and \emph{active} modes. Then the \emph{duty cycle} is defined as the percentage of time for the active mode.  In the silent mode, the  energy of the incident wave is fully harvested without reflection, by matching  the variable impedance to that of the antenna, and the circuit may be turned off for energy conservation. In the active mode, the Tag circuit is activated to receive or transmit data by backscattering where only the fraction of harvested energy is much smaller (see Fig.~\ref{fig:architecture}). Consequently, the duty-cycle is a key design parameter for regulating an energy-rate tradeoff.
{Also, we refer the interested reader to a recent survey paper, i.e., \cite{AmBackSurvey}, with detailed comparison of different BackCom systems including energy sources, operating range, bandwidths, multiple access schemes.}

\section{Next Generation BackCom System and Network Architectures}
%{\color{red} Rewrite following the outline below. The original Reader-Tag configuration can be mentioned in the text but no need to have separate sub-section
%\begin{itemize}
%\item Multiple-access BackCom systems
%\item BackCom Systems with Ambient Energy Harvesting
%\item Wirelessly Powered  BackCom Systems
%\item BackCom Systems with Technology Conversion
%\end{itemize}
%
%As there are overlapping with next section, please merge Section V into this section.
%}

\begin{figure}
	\centering
	\resizebox{0.4\textwidth}{!}{%
		\includegraphics[scale=0.82]{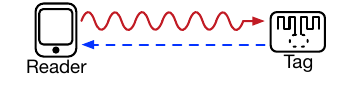}	
	}%
	\caption{Conventional BackCom.}
	\label{fig:networks_1}
\end{figure}

{The conventional BackCom system discussed in Sec.~\ref{sec:basic} is the simplest BackCom system, as illustrated in Fig.~\ref{fig:networks_1}.
The next generation BackCom systems for IoT are much more complex and can be divided into the following categories.}

\subsection{Multiple-Access BackCom Systems}
\begin{figure}
	\centering
	\resizebox{0.4\textwidth}{!}{%
		\includegraphics[scale=0.82]{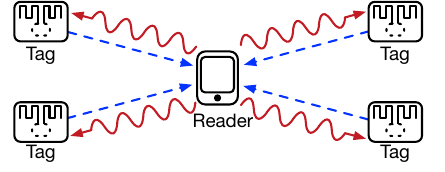}	
	}%
	\caption{Multiple-Access BackCom.}
	\label{fig:networks_2}
\end{figure}

Many real-life applications can be modeled as a {multiple-access} (MAC) BackCom system where a single Reader serves multiple Tags, as illustrated in Fig.~\ref{fig:networks_2}.
For instance, in a warehouse, an administrator can use a single Reader to collect  information simultaneously  from hundreds or thousands of items equipped with RFID Tags.
In a smart city, a data aggregator can receive sensing data from a large number of backscatter sensors at the same time.

{The key challenge in multiple-access BackCom systems is how to deal with collisions that arise as a result of concurrent Tag transmissions. In this regard, a simple solution is to avoid collisions between multi-Tag transmissions} using the traditional MAC schemes including {space/frequency/code/time-division multiple-access} (SDMA/FDMA/CDMA/TDMA), see~\cite{Boyer14} and~\cite{Bletsas} and references therein.
%The other scheme, {orthogonal frequency-division multiple-access} (OFDMA),  may not be suitable as  the required  FFT operation  is too complex for backscatter nodes.
{In SDMA, the Reader scans the space around it using directional antennas. The Tags in the reading range are distinguished by their angular position.
	Therefore, the SDMA scheme requires large antenna arrays at the Readers which increases the complexity.
In FDMA, the Tags adjust the frequency of alternation between
two switch states on the order of Hz or kHz, and the Reader detects Tag's signal in frequency domain.
Thus, the FDMA scheme incurs high signal processing complexity due to the fast Fourier transform (FFT) computation.
In CDMA, different Tags use different orthogonal codes to modulate their signals. Due to the near-far problem, the Tags of a CDMA network are required to do power control (i.e., to adjust the reflection coefficients) so as to achieve the same power level at the Reader. This introduces higher protocol complexity.
Due to its simplicity, TDMA is perhaps the most practical scheme for MAC BackCom systems where Tags transmit in separate  pre-assigned time slots in each frame. The inherent closed-loop signaling for BackCom facilitates the needed Reader-Tag synchronization for TDMA.}
Researchers have also attempted to design new MAC schemes exploiting the characteristics of BackCom. For instance,  an interesting new method for collision avoidance was proposed  in~\cite{Katabi12} for MAC BackCom that  treats bursty backscatter transmissions by Tags as a sparse code and decodes multi-Tag data at the Reader  using a compressive-sensing algorithm.

\subsection{Ad Hoc BackCom Systems} \label{subsec:interference}
\begin{figure}
	\centering
	\resizebox{0.4\textwidth}{!}{%
		\includegraphics[scale=0.82]{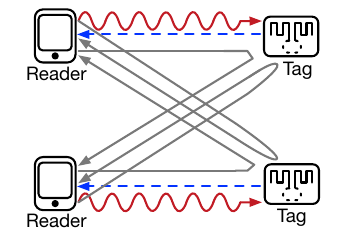}	
	}%
	\caption{BackCom Interference Network.}
	\label{fig:networks_3}
\end{figure}

{To  avoid unnecessarily  overloading the core network and to reduce latency, distributed {device-to-device} or ad hoc communications is envisaged in future IoT, creating BackCom interference channels.}
Compared with conventional interference channels, a backscatter node reflects all incident interference signals, resulting in \emph{interference regeneration}. The phenomenon is illustrated in Fig.~\ref{fig:networks_3} showing a two-link BackCom interference channel where Readers 1 and 2 each transmit a CW to backscatter Tags~1 and~2, respectively. In total, each Reader, e.g. Reader~1, is exposed to two interference components  due to Tag 2's modulation and reflection of CWs from Readers~1 and~2. In theory, the number of interference components received by each Reader can increase as the square of the number of coexisting links instead of linearly with the number  in the conventional case. As a result,  the interference issue is particularly severe in BackCom interference networks due to interference regeneration. One effective way for coping with this issue is to adopt \emph{spread spectrum} techniques in BackCom exploiting its low data rates~\cite{WanchunBackCom}.
%Among others, perhaps the most energy efficient technique is time-hopping spread spectrum since signals are sparse and allow relatively low energy consumption by the Tags~\cite{WanchunBackCom}.

\subsection{Power Beacon and Ambient RF Powered BackCom Systems}
\begin{figure}
	\centering
	\resizebox{0.4\textwidth}{!}{%
		\includegraphics[scale=0.82]{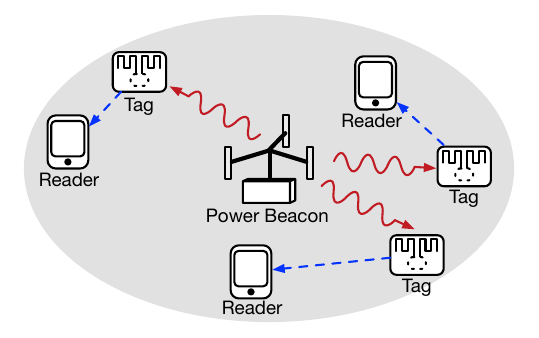}	
	}%
	\caption{Wirelessly Powered BackCom.}
	\label{fig:networks_4}
\end{figure}

{In the future IoT, most transmitting nodes are expected to be energy constrained and cannot act as Readers to emit high-power CW and power their receivers nor to enable Reader-to-Tag transmissions.} For IoT, there are two practical solutions for the power-source problem:

One solution is to deploy dense low-complexity and low-cost {power beacons} (PBs) dedicated for microwave power transfer (MPT)~\cite{Huang15}, to enable  BackCom links in their proximity by beaming towards them strong CWs, as illustrated in Fig.~\ref{fig:networks_4}. Then the Tag is able to perform RF energy harvesting and BackCom to the Reader.
Such BackCom system is termed as \emph{wirelessly powered BackCom}.

\begin{figure}
	\centering
	\resizebox{0.4\textwidth}{!}{%
		\includegraphics[scale=0.82]{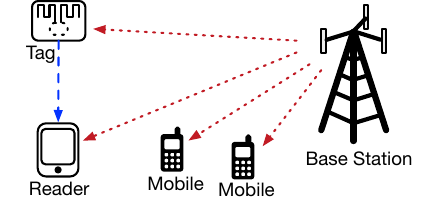}	
	}%
	\caption{Ambient Energy-Harvesting BackCom.}
	\label{fig:networks_5}
\end{figure}

The other solution is to harvest the energy from ambient RF signals, such as the signals from cellular, TV broadcasting and WiFi networks.
Leveraging the ambient RF signals can allow direct Tag-to-Reader communication  without Readers supplying power, which has motivated various relevant designs recently~\cite{Ambient13}. For instance, a backscatter Tag can transmit data to a peer by reflecting an incident base-station signal, as illustrated in Fig.~\ref{fig:networks_5}~\cite{MMAC}.
Therefore, instead of having some specific PBs or Readers that enables  tag-to-reader communications, the ambient backscatter system can use existing infrastructure and benefit from signals that are not intended for itself.
Such an energy harvesting BackCom system differs from one with a Reader or PB in two important aspects.
First, the CW is replaced with a modulated signal and thus the reflection is \emph{double modulated} with superimposed unintended and intended data. This problem can be tackled by exploiting the asymmetry in the high-rate ambient and low-rate backscatter signals. As result, the latter can be detected after suppressing the former by time averaging over each symbol duration.
The second issue is the incident ambient signal is much weaker than the CW from an intended Reader or PB due to a {much longer propagation} distance.  Consequently, BackCom with ambient RF signal is suitable only for either short-distance or infrequent Tag-to-Reader communications.

\subsection{BackCom Systems with Technology Conversion}
{To communicate with passive Tags embedded with different types of available commercialized devices,
we envisage BackCom systems with technology conversion to become common place.}
% to enable BackCom using only commodity devices

\begin{figure}
	\centering
	\resizebox{0.4\textwidth}{!}{%
		\includegraphics[scale=0.82]{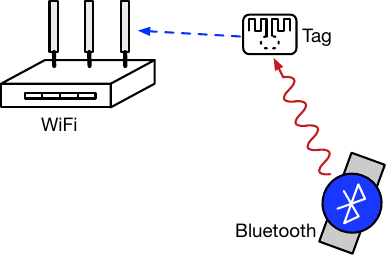}	
	}%
	\caption{BackCom System with Technology Conversion.}
	\label{fig:networks_6}
\end{figure}

One technology conversion is to leverage a Bluetooth signal for BackCom between a Tag and a WiFi device~\cite{SIGCOMM16}, as illustrated in Fig.~\ref{fig:networks_6}.
{Consider a wearable-device network, which consists of an implanted BackCom sensor (i.e., a Tag), a Bluetooth low energy (BLE) watch, and a smart phone, which is a WiFi device.
Based on the BLE protocol, the Bluetooth watch can utilize one advertisement channel of the 2.4 GHz ISM band and adopt a {Gaussian Frequency Shift Keying} (GFSK) method that encodes bits using two frequency tones~\cite{SIGCOMM16}\footnote{ Note that a typical Bluetooth device uses frequency hopping technique across the $36$ \emph{data channels} spread across the $2.4$ GHz ISM band, and uses GFSK modulation on three \emph{advertisement channels}.}.}
Hence, the watch emits a CW at either of these frequency tones.
By leveraging such CW-like Bluetooth signals, the BackCom Tag can get its sensed e-health information collected by the smart phone through backscatter modulation.
Although the Bluetooth CW frequency is usually different from WiFi, by adopting a FSK modulation through properly switching between different impedances, the BackCom Tag is able to shift the Bluetooth CW to the WiFi channel, hence achieving the BackCom with the smart phone.

Another technology conversion is to leverage WiFi. Consider a BackCom system for smart home applications, which consists of a WiFi {access point} (AP) and a backscatter IoT sensor.
The AP transmits packets to the WiFi clients, such as laptops and smart phones, which is also received by the backscatter IoT sensor.
Then the sensor is able to modulate its data on the unintended signal and backscatters the signal to the AP.
The double modulated backscattered signal is used by the AP, which obtains the Tag's information by removing its own transmitted information.
In this way, sensors can be powered by WiFi APs and also rely on them to access the Internet even in the presence of access by WiFi devices, thereby providing inter-connectivity of everything for smart homes.

\subsection{Comparison with a Traditional System}
{{In order to show the advantage of the next generation BackCom systems over traditional wireless sensor networks (WSN), we consider the example of a wirelessly powered BackCom system.
Using MATLAB, we focus on a system-level simulation of an IoT network,} where each IoT sensor node adopts either a backscatter circuit or a traditional RF circuit (including a mixer, DAC, and amplifier).
Both types of the nodes are powered by a PB.
The system level application of the simulation can be a smart supermarket.
For example, the customers in a shopping mall/supermarket may use their smart phones (which act as Readers) to collect information from different goods on the shelf (which are sent by Tags).

	\begin{table*}
	\small
	\caption{{Comparison of Active Components of BackCom and Traditional IoT Nodes.}}
	\label{tab:table2}
{	
	\hspace{-1.5cm}
	\begin{tabular}{ l | c  | c| c| c| c |c }
		\hline			
		& Sensor & MCU  & Variable Impedance & DAC & Mixer & Power Amplifier\\
		\hline
		\begin{tabular}{l}
			Energy/Power\\Consumption
		\end{tabular}
		&
		\begin{tabular}{l}
			$0.1~\mu$J per \\sensing task
		\end{tabular}
		& $2.5~\mu$W  & $0.1~\mu$W & $15~\mu$W & $0.1$~mW  & $50$~mW \\
		\hline
		\hline
		BackCom IoT Node & \checkmark  & \checkmark  & \checkmark &  &   &    \\
		\hline
		Traditional IoT Node & \checkmark & \checkmark  &  &\checkmark & \checkmark & \checkmark  \\
		\hline			
		%			\hspace{-0.22cm}\begin{tabular}{l}
		%				Next Generation \\BackCom
		%			\end{tabular} & $< 3$ km & $< 10$ Mbps  & \begin{tabular}{l}
		%				Higher-order\\ modulation
		%			\end{tabular} & \begin{tabular}{l}
		%				Multiple access \\and ad hoc
		%			\end{tabular} & Dedicated and ambient  \\
		%			\hline
	\end{tabular}}
\end{table*}

The IoT nodes with a density of $0.2$ nodes/m$^2$ are randomly distributed in a disk region with the radius of $5$~m and served by the PB at the center.
Each node aims to perform sensing and transmission to its intended receiver at a fixed distance of $0.5$~m.
The IoT nodes adopt a harvesting-then-sensing-and-transmission task sequence per $100$-ms time slot, where energy harvesting occupies $80$~ms and the other operations $20$~ms if there is sufficient energy.}

The antenna effective apertures of all nodes are $0.001$~m$^2$.
We consider Friis free-space channels for wireless power transfer and information transmission.
The carrier frequency is $2.4$~GHz, the modulation scheme BPSK, and the noise power at the receiver $-70$~dBm,
and the RF energy harvesting efficiency $50\%$.
{The IoT nodes use orthogonal Walsh-Hadamard code with length $16$ for information transmission.}
{Note that each communication pair has no information of the other pairs and do not perform power control.}
Each sensing task consumes $0.1$~$\mu$J of energy (e.g., ambient light sensor TSL2550D), and the digital circuit power consumption during the active mode is $2.5$~$\mu$W~\cite{Tagconsumption}.
Each BackCom IoT node does not perform RF energy harvesting while backscattering. {While backscattering using BPSK modulation, the Tag switches the value of the load impedance between zero (a short state) and infinite (an open state).}
{The impedance switching frequency is equal to the chip rate of the orthogonal code.
	The switch (e.g., DG2012~\cite{CMOS}) has typical power consumption of around $0.1~\mu$W.}
{For the traditional IoT node, the power consumptions of the DAC and the mixer during active mode are $15$~$\mu$W and $0.1$~mW, respectively (e.g.,  DAC8830 and AD831), and the power amplifier (e.g., LMH6609) has power consumption of about $50$~mW.
	Note that since the local oscillator (LO) is the most important element of the mixer, the power consumption of the mixer is approximated by that of the LO.
The active components and their energy/power consumption are listed in Table~\ref{tab:table2}.}

\begin{figure}[t]
	\small
	\centering
	%	\vspace*{-0.7cm}	
	\resizebox{0.4\textwidth}{!}
	{\includegraphics[scale=0.2]{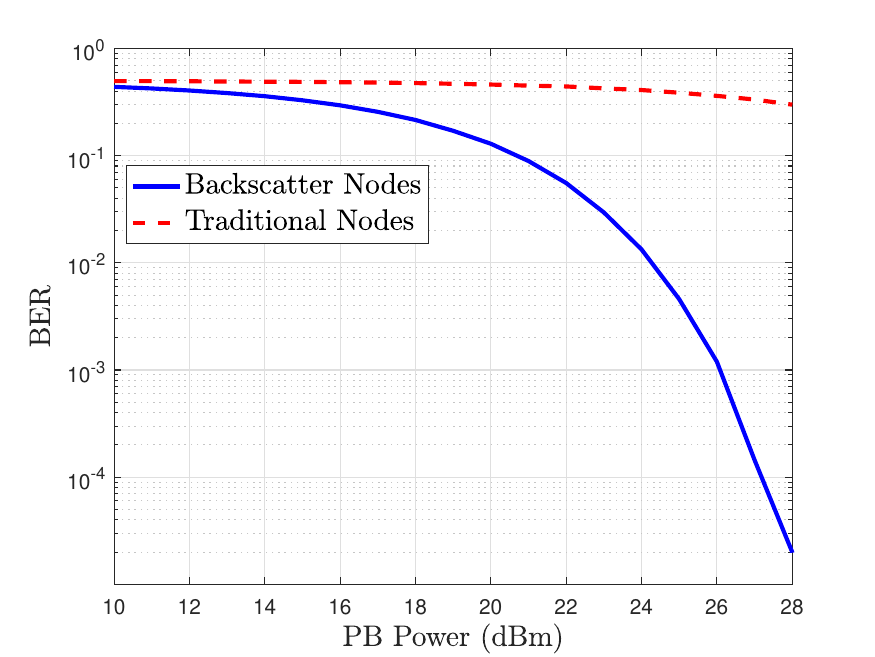}}	
	\caption{\small BER versus PB power.}	
	\label{fig:BER}
\end{figure}

\begin{figure}[t]
	\small
	\centering
	%	\vspace*{-0.7cm}	
	\resizebox{0.4\textwidth}{!}
	{\includegraphics[scale=0.2]{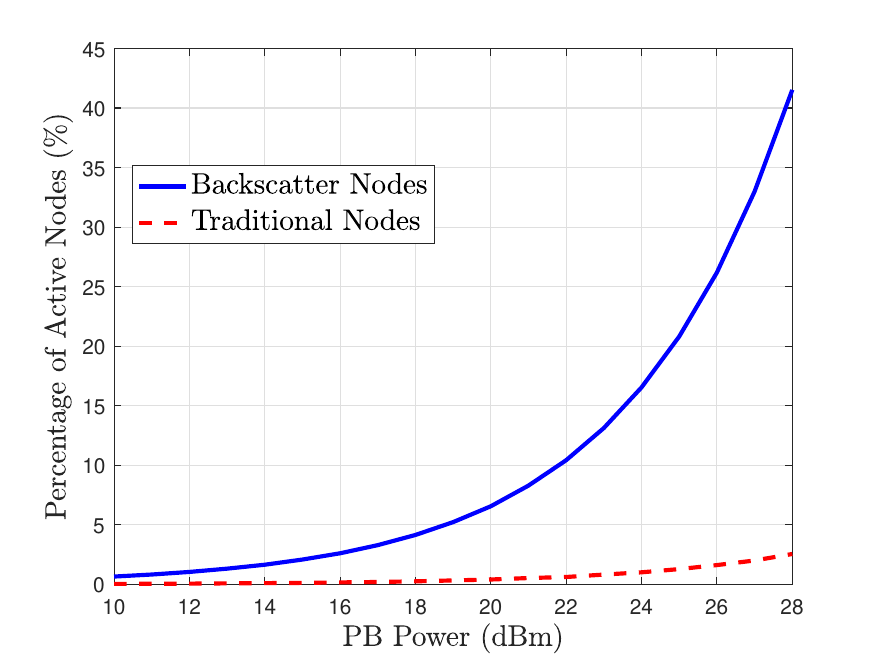}}	
	\caption{\small Percentage of active nodes versus PB power.}	
	\label{fig:active}
\end{figure}

{Based on these practical settings, we investigate how much performance improvement is achieved in the IoT network by adopting the backscatter circuit.
Fig.~\ref{fig:BER} plots the average BER at the receiver versus the PB's transmit power.
We can see that the BER achieved by the backscatter nodes is significantly reduced compared with the traditional nodes within a practical range of PB's transmit power, e.g., the BER is reduced by $100$ times when the PB has a $25$~dBm transmission power\footnote{This PB transmit power is practical. This is because according to FCC regulations, the transmission power should be less than 30 dBm for omnidirectional transmission.}.
Fig.~\ref{fig:active} plots the percentage of active nodes.
We see that the percentage of active IoT nodes increases with the PB's transmit power, and
the percentage of the active backscatter nodes is much larger than that of the traditional nodes.
For example, the improvement of the percentage of active nodes is $7\%$ when the transmit power is $20$~dBm, and the improvement is $20\%$ when the transmit power is $25$~dBm.
Therefore, the simulation results quantitatively show that the backscatter nodes can significantly improve the performance of the IoT network, and the wirelessly powered BackCom is a promising solution for future IoT applications. In the following section, we present more advanced BackCom techniques.}

\section{Emerging BackCom Techniques}
The future IoT applications require the BackCom systems to enable long-distance, low-latency, and high-rate communications between massive IoT devices and also enable their ad hoc communication.
A few advanced BackCom techniques aiming at tackling these challenges are discussed as follows.

\subsection{BackCom Systems with Power Beacons}
Different from conventional RFID Tags which only need to report its ID information, the IoT BackCom Tags are also required to perform sensing and computing which can consume more energy.
How to wirelessly power BackCom using PBs (see earlier discussions) and maximize the power-transfer efficiency is a significant design issue.

The efficiency can be increased by energy beamforming from a multi-antenna PB and a Tag.
To this end, the PB needs to know the forward {channel state information} (CSI).
For a free-space channel, the CSI reduces to the Tag direction with respect to the PB.
The PB can form a beam using the well-known retrodirective beamforming technique that automatically generates a beamformer by conjugating the observation of the pilot signal sent (i.e., reflected) by the Tag.
On the other hand, for a scattering channel, the beamformer designs are more complex but can exploit the so called ``key-hole" channel structure due to backscattering (see e.g.~\cite{Yang15}).

Another technique for improving the power-transfer efficiency is to optimize the CW waveform, e.g., as a weighted sum of multiple sinusoidal waves (see e.g.,~\cite{Bruno16}).
The design aims at increasing the peak-to-average power ratio of the PB signal that yields a higher energy-harvesting efficiency due to its non-linearity as a function of incident power.

\subsection{Full-Duplex BackCom}
For future IoT, there would be massive number of Tag-to-Reader links existing at the same time.
Though information flow in RFID applications is usually uni-directional, message exchange between nodes is common in IoT.
{Thus, full-duplex communication can substantially reduce the latency, and improve the efficiency of spectrum utilization of the IoT-Reader-to-Tag links.}
{For BackCom full-duplexing, there is a performance tradeoff between the transmission and the reception of a full-duplex BackCom node.
	A smaller backscatter coefficient reduces the reflected signal power and thereby increases the received signal power, and vice versa.}

There are two methods for implementing full-duplex BackCom.
Both require a Reader to have a full-duplex antenna allowing simultaneous transmission and reception~\cite{FullDuplexGuo}.
Consider simultaneous forward and backward information transmission in a BackCom system having one pair of Reader and Tag.
For the first method, by leveraging prior knowledge of forward information, the Reader can cancel it in the backward information transmission signal and thereby retrieve the backward information. This method supports symmetric bidirectional data rates.
The other method exploits the rate asymmetry in data transfer (Tag to Reader) and control signaling (Reader to Tag).
Specifically, the signals have different frequencies and can be decoupled by filtering or averaging, see~\cite{WanchunBackCom} and references therein.

\subsection{Time-Hopping BackCom}
As mentioned, interference in IoT networks with high node density poses a key design challenge that is exacerbated by backscattering.
For IoT devices which are sensors for smart cities and homes, the burstiness in their transmissions can be exploited for tackling interference.
A suitable transmission technique is {time-hopping spread spectrum} (TH-SS), where each Tag randomly selects one of $N$ time slots for transmitting a single symbol and the choices of different Tags are independent~\cite{WanchunBackCom}.
As a result, the number of simultaneous links is reduced by the factor $N$, called the processing gain, thereby suppressing interference.
An extreme form of TH-SS can be realized by {ultra-wideband} (UWB) transmission, where an extremely large processing gain is achieved using ultra-narrow pulses whose durations are in the order of nano-second.

\subsection{MIMO BackCom}
A key characteristic of the backward BackCom (i.e., the tag-to-reader information transmission) is the \emph{double path-loss} due to the fact that the backscatter signal received at a Reader propagates through the closed-loop channel cascading the downlink and uplink channels. The resultant path loss is especially server as the propagation distances in IoT are much longer than those for RFID applications. To enhance link reliability, one solution is to deploy antenna arrays at Readers and Tags and apply spatial-diversity techniques. Furthermore, multi-antennas can enhance the efficiency of wireless power transfer by enabling transmit energy beamforming and increasing receive antenna apertures.

Backscattering introduces a special channel structure for the backward information transmission in a {multiple-input-multiple-output} (MIMO) BackCom system, called a \emph{dyadic MIMO channel}, which captures the composite fading in the forward and backward channels~\cite{Boyer14}. To be specific, the CW signals sent by the transmit antennas of the Reader propagate through the forward MIMO channel, and are first combined at each antenna of the Tag and then backscattered, and lastly propagate through the backward MIMO channel to arrive
at the receive antennas of the Reader. The resultant dyadic MIMO channel has a similar structure as the classic \emph{keyhole MIMO channel}. The space-time coding is a simple but suitable technique for achieving the diversity gain of such a channel. By adopting space-time coding, it has been proved that the achievable maximum diversity order is equal to the number of the Tag's antennas~\cite{Boyer14}. In other words, in contrast to the conventional MIMO channel, increasing the number of receive antenna at the Reader cannot continuously enhance the reliability of the backward information transmission.

\section{IoT Applications for BackCom}

\subsection{BackCom for Smart Homes/Cities}

Low-power or passive BackCom devices with energy harvesting capabilities can be densely deployed to provide pervasive and uninterrupted sensing and computing services that provide a platform for implementing applications for smart homes/cities.

\begin{figure}[t]
	\small
	\centering
	%	\vspace*{-0.7cm}	
	\resizebox{0.4\textwidth}{!}
	{\includegraphics[scale=0.2]{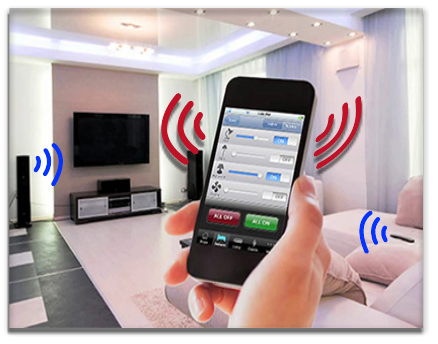}}	
	\caption{\small BackCom application: smart home.}	
	\label{fig:app_1}
\end{figure}

In a smart home, a large number of passive BackCom sensors can be placed at flexible locations (e.g., embedded in walls, ceilings, and furniture). They are freed from the constraints due to recharging or battery replacements as one or multiple in-house PBs can be deployed to simultaneously power all the sensors or otherwise they can operate on ambient energy harvesting. The tasks performed by the sensors have a wide range such as detection of gas leak, smoke and CO, monitoring movements, indoor positioning, and surveillance [see Fig.~\ref{fig:app_1}].  As an example, BackCom-based smart dustbins are able to monitor their trash levels and communicate the information to passing-by garbage trucks by backscattering, streamlining the trash-collection process. Another example is that household robots are able to use the backscattered signals from the Tags located on doors and furniture for indoor navigation~\cite{RFIDRobot}.

In a smart city, ubiquitous BackCom sensor nodes can be placed in every city corner such as buildings, bridges, trees, street lamps, and parking areas. They can streamline the city operations and improve our life quality via e.g., monitoring of air/noise pollution and traffic and parking-availability indicating. The efficient sensing-data fusion and wireless power for BackCom sensors can be realized by the deployment of integrated PBs and APs at fixed locations or mounted on autonomous ground vehicles or UAVs, providing full-city coverage without costly backhaul networks \cite{YongZeng16}.

\begin{figure}[t]
	\small
	\centering
	%	\vspace*{-0.7cm}	
	\resizebox{0.4\textwidth}{!}
	{\includegraphics[scale=0.2]{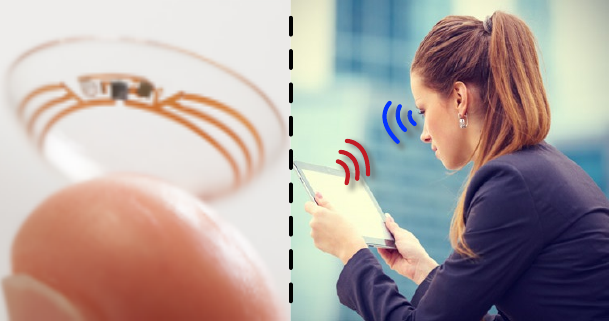}}	
	\caption{\small BackCom application: smart contact lens.}	
	\label{fig:app_2}
\end{figure}

\begin{figure}[t]
	\small
	\centering
	%	\vspace*{-0.7cm}	
	\resizebox{0.4\textwidth}{!}
	{\includegraphics[scale=0.2]{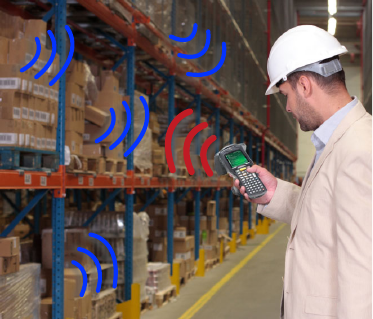}}	
	\caption{\small BackCom application: logistics management.}	
	\label{fig:app_3}
\end{figure}

\subsection{BackCom for Biomedical Applications}

IoT biomedical applications, such as plant/animal monitoring, wearable, and implantable human health monitoring, require tiny and low heat-radiation communication devices. BackCom devices, which do not rely on any active RF component, can meet such requirements and thereby avoid causing any significant effect on the plants, animals, tissues or organs being monitored. These advantages make BackCom a promising solution for IoT biomedical applications. One example is the  BackCom-based smart Google Contact Lens, as illustrated in Fig.~\ref{fig:app_2}. The lens was invented in Google in 2014 for the purpose of assisting people with diabetes by constantly measuring the glucose levels in their tears (once per second). The device consists of a miniaturized glucose sensor and a tiny BackCom Tag. The Tag is able to provide energy to the sensor by RF energy harvesting from a wireless controller, and also backscatter the measured blood sugar level to the wireless controller for diagnosing purpose. Looking into the future, we envisage that BackCom will find a wide range of biomedical applications. In particular, implantable tiny BackCom neural devices with ultra-low power consumption and heat radiation may be placed on the surface of the patient's brain to help the study, diagnosis and treatment of diseases such as epilepsy and Parkinson's disease, where the BackCom implants act as the brain-computer interface.

\subsection{BackCom for Logistics}
BackCom for logistics is a very attractive proposition due to the ultra-low manufacturing cost of simple and passive BackCom Tags, as illustrated in Fig.~\ref{fig:app_3}. For example, as early as 2007, the biggest $100$ suppliers of the global renowned chain commercial group Wal-Mart have used the BackCom technology for logistics tracking. The technology has been helping the companies to substantially reduce operational cost, guarantee product quality, and accelerate the processing speed. In the past decade, the popularization and the application of BackCom have brought revolutionary changes to the logistics industry, due to its advantages compared with the conventional bar code technology, such as reduced manual control, long service lives, long reading distances, and encrypt-able and rewritable data.

Looking into the future, apart from the existing BackCom techniques for logistics tracking and management, BackCom-based three-dimensional orientation tracking is an emerging technique. By attaching an array of low-cost passive BackCom Tags as orientation sensors on the surface of the target objects, three-dimensional orientation information is available at the Reader by analyzing the relative phase offset between different Tags. In this way, human workers can be warned when the angle of a cargo is larger than a threshold.

\section{Concluding Remarks}
IoT that aims to enable both the activity and the connectivity of billions of energy-consuming smart nodes, is challenged from the energy perspective. The BackCom systems and techniques provide promising solutions for tackling this challenge.
In this article, we have introduced the basic principles for BackCom, summarized existing BackCom system and network architectures and discussed several emerging advanced BackCom techniques. Moreover, we have described various applications of BackCom in IoT applications. With rapid advancements in both theory and applications, the technology is expected to play a key role in future IoT by enabling truly ubiquitous network connectivity, and pervasive sensing and computing.

\begin{backmatter}
\section*{List of Abbreviations}
AP: access point;
BackCom: backscatter communications;
BLE: Bluetooth low energy;
BPSK: binary phase-shift keying;
CDMA: code-division multiple-access;
CSI: channel state information;
CW: continuous wave;
FDMA: frequency-division multiple-access;
GFSK: Gaussian frequency shit keying;
MAC: multiple access;
OFDMA: orthogonal frequency-division multiple-access;
PB: power beacon;
PT: power transfer;
RF: radio-frequency;
RFID: radio-frequency identification;
SDMA: space-division multiple-access;
TDMA: time-division multiple-access;
TH-SS: time-hopping spread spectrum;
UAV: unmanned aerial vehicle;
UWB: ultra-wideband;
WSN: wireless sensor networks

\section*{Competing interests}
The authors declare that they have no competing interests.

\section*{Funding}
The work of W. Liu was supported by the Australian Research Council's Australian Laureate Fellowships scheme (project number FL160100032).
The work of K. Huang was supported by Hong Kong Research Grants Council under the Grants 17209917 and 17259416.
The work of S. Durrani and X. Zhou was supported by the Australian Research Council's Discovery Project Funding Scheme (Project number DP170100939).

\section*{Author's contributions}
K. Huang and W. Liu came up with the original idea.
W. Liu drafted the manuscript under the supervision of K. Huang.
X. Zhou and S. Durrani helped to further refine the manuscript.
All authors read and approved the final manuscript.

\bibliographystyle{bmc-mathphys}
%\bibliography{article_cite}

\newcommand{\BMCxmlcomment}[1]{}

\BMCxmlcomment{
	
	<refgrp>
	
	<bibl id="B1">
	<title><p>Energy Harvesting Sensor Nodes: {Survey} and
	Implications</p></title>
	<aug>
	<au><snm>Sudevalayam</snm><fnm>S.</fnm></au>
	<au><snm>Kulkarni</snm><fnm>P.</fnm></au>
	</aug>
	<source>IEEE Commun. Surveys Tuts.</source>
	<pubdate>2011</pubdate>
	<volume>13</volume>
	<issue>3</issue>
	<fpage>443</fpage>
	<lpage>461</lpage>
	</bibl>
	
	<bibl id="B2">
	<title><p>A wireless sensing platform utilizing ambient {RF}
	energy</p></title>
	<aug>
	<au><snm>Parks</snm><fnm>AN</fnm></au>
	<au><snm>Sample</snm><fnm>AP</fnm></au>
	<au><snm>Zhao</snm><fnm>Y</fnm></au>
	<au><snm>Smith</snm><fnm>JR</fnm></au>
	</aug>
	<source>Proc. IEEE PAWR</source>
	<pubdate>2013</pubdate>
	<fpage>160</fpage>
	<lpage>-162</lpage>
	</bibl>
	
	<bibl id="B3">
	<title><p>A battery-less, energy harvesting device for long range scavenging
	of wireless power from terrestrial {TV} broadcasts</p></title>
	<aug>
	<au><snm>Vyas</snm><fnm>R</fnm></au>
	<au><snm>Nishimoto</snm><fnm>H</fnm></au>
	<au><snm>Tentzeris</snm><fnm>M</fnm></au>
	<au><snm>Kawahara</snm><fnm>Y</fnm></au>
	<au><snm>Asami</snm><fnm>T</fnm></au>
	</aug>
	<source>Proc. IEEE MTT-S</source>
	<pubdate>2012</pubdate>
	<fpage>1</fpage>
	<lpage>-3</lpage>
	</bibl>
	
	<bibl id="B4">
	<title><p>PowerSpot</p></title>
	<pubdate>2017</pubdate>
	<url>https://www.powercastco.com/wp-content/uploads/2018/06/TX91503-UserManualREV4.pdf</url>
	</bibl>
	
	<bibl id="B5">
	<title><p>Cutting the last wires for mobile communications by microwave power
	transfer</p></title>
	<aug>
	<au><snm>Huang</snm><fnm>K</fnm></au>
	<au><snm>Zhou</snm><fnm>X</fnm></au>
	</aug>
	<source>IEEE Communications Magazine</source>
	<pubdate>2015</pubdate>
	<volume>53</volume>
	<issue>6</issue>
	<fpage>86</fpage>
	<lpage>-93</lpage>
	</bibl>
	
	<bibl id="B6">
	<title><p>Backscatter Communication and {RFID}: Coding, Energy, and {MIMO}
	Analysis</p></title>
	<aug>
	<au><snm>Boyer</snm><fnm>C</fnm></au>
	<au><snm>Roy</snm><fnm>S</fnm></au>
	</aug>
	<source>IEEE Transactions on Communications</source>
	<pubdate>2014</pubdate>
	<volume>62</volume>
	<issue>3</issue>
	<fpage>770</fpage>
	<lpage>785</lpage>
	</bibl>
	
	<bibl id="B7">
	<title><p>CC2520 DATASHEET</p></title>
	<pubdate>2007</pubdate>
	<url>http://www.ti.com/lit/ds/symlink/cc2520.pdf</url>
	</bibl>
	
	<bibl id="B8">
	<title><p>Fully integrated passive {UHF} {RFID} transponder {IC} with
	16.7-$\mu$W minimum {RF} input power</p></title>
	<aug>
	<au><snm>Karthaus</snm><fnm>U</fnm></au>
	<au><snm>Fischer</snm><fnm>M</fnm></au>
	</aug>
	<source>IEEE Journal of Solid-State Circuits</source>
	<pubdate>2003</pubdate>
	<volume>38</volume>
	<issue>10</issue>
	<fpage>1602</fpage>
	<lpage>-1608</lpage>
	</bibl>
	
	<bibl id="B9">
	<title><p>Full-Duplex Backscatter Interference Networks Based on Time-Hopping
	Spreading Spectrum</p></title>
	<aug>
	<au><snm>{Liu}</snm><fnm>W.</fnm></au>
	<au><snm>{Huang}</snm><fnm>K.</fnm></au>
	<au><snm>{Zhou}</snm><fnm>X.</fnm></au>
	<au><snm>{Durrani}</snm><fnm>S.</fnm></au>
	</aug>
	<source>IEEE Transactions on Wireless Communications</source>
	<pubdate>2017</pubdate>
	<volume>16</volume>
	<issue>7</issue>
	<fpage>4361</fpage>
	<lpage>4377</lpage>
	</bibl>
	
	<bibl id="B10">
	<title><p>DG2012 DATASHEET</p></title>
	<pubdate>2007</pubdate>
	<url>http://www.vishay.com/docs/72176/dg2012.pdf</url>
	</bibl>
	
	<bibl id="B11">
	<title><p>Ambient Backscatter Communications: A Contemporary
	Survey</p></title>
	<aug>
	<au><snm>Huynh</snm><fnm>NV</fnm></au>
	<au><snm>Hoang</snm><fnm>DT</fnm></au>
	<au><snm>Lu</snm><fnm>X</fnm></au>
	<au><snm>Niyato</snm><fnm>D</fnm></au>
	<au><snm>Wang</snm><fnm>P</fnm></au>
	<au><snm>Kim</snm><fnm>DI</fnm></au>
	</aug>
	<source>IEEE Communications Surveys \& Tutorials</source>
	<pubdate>2018</pubdate>
	<volume>20</volume>
	<issue>4</issue>
	<fpage>2889</fpage>
	<lpage>2922</lpage>
	</bibl>
	
	<bibl id="B12">
	<title><p>Anti-collision backscatter sensor networks</p></title>
	<aug>
	<au><snm>Bletsas</snm><fnm>A.</fnm></au>
	<au><snm>Siachalou</snm><fnm>S.</fnm></au>
	<au><snm>Sahalos</snm><fnm>J. N.</fnm></au>
	</aug>
	<source>IEEE Trans. Wireless Commun.</source>
	<pubdate>2009</pubdate>
	<volume>8</volume>
	<issue>10</issue>
	<fpage>5018</fpage>
	<lpage>5029</lpage>
	</bibl>
	
	<bibl id="B13">
	<title><p>Efficient and reliable low-power backscatter networks</p></title>
	<aug>
	<au><snm>Wang</snm><fnm>J</fnm></au>
	<au><snm>Hassanieh</snm><fnm>H</fnm></au>
	<au><snm>Katabi</snm><fnm>D</fnm></au>
	<au><snm>Indyk</snm><fnm>P</fnm></au>
	</aug>
	<source>Proc. ACM SIGCOMM</source>
	<pubdate>2012</pubdate>
	<fpage>61</fpage>
	<lpage>-72</lpage>
	</bibl>
	
	<bibl id="B14">
	<title><p>Ambient backscatter: {Wireless} communication out of thin
	air</p></title>
	<aug>
	<au><snm>Liu</snm><fnm>V</fnm></au>
	<au><snm>Parks</snm><fnm>A</fnm></au>
	<au><snm>Talla</snm><fnm>V</fnm></au>
	<au><snm>Gollakota</snm><fnm>S</fnm></au>
	<au><snm>Wetherall</snm><fnm>D</fnm></au>
	<au><snm>Smith</snm><fnm>JR</fnm></au>
	</aug>
	<source>ACM SIGCOMM Computer Communication Review</source>
	<pubdate>2013</pubdate>
	<volume>43</volume>
	<issue>4</issue>
	<fpage>39</fpage>
	<lpage>-50</lpage>
	</bibl>
	
	<bibl id="B15">
	<title><p>Backscatter Multiplicative Multiple-Access Systems: Fundamental
	Limits and Practical Design</p></title>
	<aug>
	<au><snm>Liu</snm><fnm>W.</fnm></au>
	<au><snm>Liang</snm><fnm>Y.C.</fnm></au>
	<au><snm>Li</snm><fnm>Y.</fnm></au>
	<au><snm>Vucetic</snm><fnm>B.</fnm></au>
	</aug>
	<source>IEEE Trans. Wireless Commun.</source>
	<pubdate>2018</pubdate>
	<volume>17</volume>
	<issue>9</issue>
	<fpage>5713</fpage>
	<lpage>5728</lpage>
	</bibl>
	
	<bibl id="B16">
	<title><p>Inter-Technology Backscatter: {Towards} Internet Connectivity for
	Implanted Devices</p></title>
	<aug>
	<au><snm>Iyer</snm><fnm>V</fnm></au>
	<au><snm>Talla</snm><fnm>V</fnm></au>
	<au><snm>Kellogg</snm><fnm>B</fnm></au>
	<au><snm>Gollakota</snm><fnm>S</fnm></au>
	<au><snm>Smith</snm><fnm>J</fnm></au>
	</aug>
	<source>Proc. ACM SIGCOMM</source>
	<pubdate>2016</pubdate>
	<fpage>356</fpage>
	<lpage>-369</lpage>
	</bibl>
	
	<bibl id="B17">
	<title><p>Multi-antenna wireless energy transfer for backscatter
	communication systems</p></title>
	<aug>
	<au><snm>Yang</snm><fnm>G</fnm></au>
	<au><snm>Ho</snm><fnm>CK</fnm></au>
	<au><snm>Guan</snm><fnm>YL</fnm></au>
	</aug>
	<source>IEEE Journal on Selected Areas in Communications</source>
	<pubdate>2015</pubdate>
	<volume>33</volume>
	<issue>12</issue>
	<fpage>2974</fpage>
	<lpage>-2987</lpage>
	</bibl>
	
	<bibl id="B18">
	<title><p>Waveform Design for Wireless Power Transfer</p></title>
	<aug>
	<au><snm>Clerckx</snm><fnm>B.</fnm></au>
	<au><snm>Bayguzina</snm><fnm>E.</fnm></au>
	</aug>
	<source>IEEE Transactions on Signal Processing</source>
	<pubdate>2016</pubdate>
	<volume>64</volume>
	<issue>23</issue>
	<fpage>6313</fpage>
	<lpage>6328</lpage>
	</bibl>
	
	<bibl id="B19">
	<title><p>In-Band Full-Duplex Wireless: Challenges and
	Opportunities</p></title>
	<aug>
	<au><snm>Sabharwal</snm><fnm>A.</fnm></au>
	<au><snm>Schniter</snm><fnm>P.</fnm></au>
	<au><snm>Guo</snm><fnm>D.</fnm></au>
	<au><snm>Bliss</snm><fnm>D. W.</fnm></au>
	<au><snm>Rangarajan</snm><fnm>S.</fnm></au>
	<au><snm>Wichman</snm><fnm>R.</fnm></au>
	</aug>
	<source>IEEE J. Sel. Areas Commun.</source>
	<pubdate>2014</pubdate>
	<volume>32</volume>
	<issue>9</issue>
	<fpage>1637</fpage>
	<lpage>1652</lpage>
	</bibl>
	
	<bibl id="B20">
	<title><p>Accurate Self-Localization in {RFID} Tag Information Grids Using
	{FIR} Filtering</p></title>
	<aug>
	<au><snm>Pomárico Franquiz</snm><fnm>J. J.</fnm></au>
	<au><snm>Shmaliy</snm><fnm>Y. S.</fnm></au>
	</aug>
	<source>IEEE Trans. Ind. Informat.</source>
	<pubdate>2014</pubdate>
	<volume>10</volume>
	<issue>2</issue>
	<fpage>1317</fpage>
	<lpage>1326</lpage>
	</bibl>
	
	<bibl id="B21">
	<title><p>Wireless communications with unmanned aerial vehicles:
	{Opportunities} and challenges</p></title>
	<aug>
	<au><snm>Zeng</snm><fnm>Y.</fnm></au>
	<au><snm>Zhang</snm><fnm>R.</fnm></au>
	<au><snm>Lim</snm><fnm>T. J.</fnm></au>
	</aug>
	<source>IEEE Communications Magazine</source>
	<pubdate>2016</pubdate>
	<volume>54</volume>
	<issue>5</issue>
	<fpage>36</fpage>
	<lpage>42</lpage>
	</bibl>
	
	</refgrp>
} % end of \BMCxmlcomment

\section*{Figures}
\begin{enumerate}
	\item {\textbf{Figure 1.} The architecture of a backscatter Tag.}

	\item {\textbf{Figure 2.} The forward information-transmission mode of a BackCom link.}

	\item {\textbf{Figure 3.} The backward information-transmission mode of a BackCom link.}

	\item {\textbf{Figure 4.} Conventional BackCom.}

	\item {\textbf{Figure 5.} Multiple-Access BackCom.}
	
	\item {\textbf{Figure 6.} BackCom Interference Network.}

	\item {\textbf{Figure 7.} Wirelessly Powered BackCom.}

	\item {\textbf{Figure 8.} Ambient Energy-Harvesting BackCom.}

	\item {\textbf{Figure 9.} BackCom System with Technology Conversion.}
	
	\item {\textbf{Figure 10.} BER versus PB power.}

	\item {\textbf{Figure 11.} Percentage of active nodes versus PB power.}

\item {\textbf{Figure 12.} BackCom application: smart home.}

\item {\textbf{Figure 13.} BackCom application: smart contact lens.}

\item {\textbf{Figure 14.} BackCom application: logistics management.}

\end{enumerate}

\end{backmatter}

\end{document}